\documentclass[twocolumn,prd,floatfix]{revtex4}
\usepackage{color}
\usepackage{tabularx}
\usepackage{epsfig}
\usepackage{amsmath}
\usepackage{amssymb}
\usepackage{bm}
\usepackage{graphicx}
\usepackage{multirow}

\usepackage{epsfig}

\begin{document}

\title{Horizon Entropy Refined: Quantum Contributions and Cosmological Insights}

\author{Alireza Maleki$^1$, Ahmad Sheykhi$^2$}
\affiliation {$^1$Department of Physics, Sharif University of Technology, Tehran, Iran
\\
$^2$Department of Physics, College of Science, Shiraz University, Shiraz 71454, Iran
}

\date{\today}
\begin{abstract}
We study the effects of quantum fluctuations on the event
horizon area and their implications for corrections to the
Bekenstein-Hawking entropy. These quantum corrections are
incorporated into the framework of large-scale gravitational
systems, utilizing the holographic principle to derive modified
Friedmann equations. By redefining the Bekenstein-Hawking entropy,
our model predicts significant alterations to the Friedmann
equations within specific parameter ranges, offering novel 
perspectives on cosmological scales. Using distance modulus data
from the Pantheon supernova sample, we demonstrate the model's
potential to constrain the parameters governing quantum
corrections and address unresolved cosmological issues. Crucially,
our analysis reveals that quantum fluctuations can increase the
area of the event horizon by up to 47\%. Beyond this threshold,
theoretical predictions encounter substantial challenges when
compared with observational data. This approach bridges quantum
gravity and observational cosmology, opening new avenues for
testing and refining theoretical models.

\end{abstract}

\maketitle

\section{Introduction}

Exploring the quantum nature of black holes remains a cornerstone
in addressing fundamental questions within black hole physics
\cite{susskindB2004}. While general relativity has traditionally
served as the foundation for understanding black holes, quantum
effects in the vicinity of the event horizon introduce significant
modifications to physical phenomena in these extreme regimes
\cite{calmet2014quantum,maleki2020speed}.

A particularly striking consequence of the interplay between
quantum mechanics and general relativity is the evaporation of
black holes through Hawking radiation \cite{Hawking1975}. 
However, these advances have also given rise to profound
theoretical challenges, such as the black hole information paradox
\cite{bekenstein1973black, hawking1973large, hooft1985quantum,
susskind1997black, callan1996d}. Addressing such paradoxes
requires a rigorous and careful exploration of quantum phenomena
in gravitational settings
\cite{susskindB2004,hawking2014,chen2015black,maleki2022complementarity}.
Moreover, a refined understanding of quantum processes near black
holes may pave the way toward a unified framework that seamlessly
integrates quantum mechanics with general relativity on
cosmological scales.

On the other hand, the discovery of the accelerated expansion of
the universe in 1998, driven by Type Ia Supernovae observations
\cite{riess1998observational}, marked a transformative development
in cosmology. Subsequent investigations \cite{haridasu2017strong}
have consistently reaffirmed this acceleration, which, within
Einstein’s gravitational framework, is attributed to the
cosmological constant $\Lambda$ \cite{zee2001einstein}. The
phenomenon driving this accelerated expansion, commonly referred
to as Dark Energy (DE), represents one of the most profound
mysteries in modern physics. While the cosmological constant
provides a straightforward theoretical interpretation of DE,
associating it with quantum vacuum energy, the striking
discrepancy between theoretical predictions and observed energy
scales—known as the cosmological constant problem—remains
unresolved \cite{weinberg2008cosmology}.

In parallel, the longstanding enigma of Dark Matter (DM) adds
another layer of complexity to our understanding of the universe.
Initially proposed by Zwicky in 1935 and later substantiated
through galactic rotation curves and large-scale structure
observations \cite{maleki2020constraint, weinberg2008cosmology},
DM plays a pivotal role in the standard cosmological model. The
$\Lambda$CDM framework, which combines Cold Dark Matter (CDM) with
the cosmological constant, offers a robust explanation for both
cosmic acceleration and the gravitational effects of DM. Planck
satellite data further elucidate this composition in this model,
revealing that the universe consists of approximately 68\% dark
energy, 28\% dark matter, and 4\% baryonic matter
\cite{aghanim2018planck}. Despite its successes, the $\Lambda$CDM
model encounters notable challenges, as recent observational data
reveal inconsistencies that suggest its limitations in fully
accounting for certain phenomena. To address these discrepancies,
various alternative theories have been proposed, aiming either to
refine the $\Lambda$CDM model or to introduce modifications to
Einstein’s field equations \cite{weinberg2008cosmology,
maleki2020investigation}.\\
A promising avenue for exploring the nature of DE involves the
thermodynamic interpretation of gravity, specifically through the
holographic principle \cite{susskind1995world,
wang2017holographic}. By incorporating Bekenstein-Hawking entropy
 into the study of large-scale gravitational
systems and utilizing the holographic principle, researchers have
successfully derived the Friedmann equations from a thermodynamic
perspective \cite{padmanabhan2014general, wang2006thermodynamics,
sheykhi2010entropic, sheykhi2018modified}.\\
This study introduces a modified entropy for Bekenstein-Hawking
entropy, derived from quantum fluctuation effects, and examines
its implications using cosmological observational data. This
modification alters the entropy of the event horizon, resulting in
substantial corrections to the derived Friedmann equations. These
corrections provide a novel framework for understanding the
quantum nature of gravity and its far-reaching impacts on
cosmological scales.

%%%%%%%%%%%%%%%%%%%%%%%%%%%%%%%%%%%%%%%%%%%%%%%%%%%%%%%%%%%%%%%%%%%%%%%%%%%%%%%%%%%%%%%
\section{Cosmology of Holographic Entropies}
Among various proposals, a significant hypothesis that contributes
to understanding the nature of Dark Energy (DE) is based on the
holographic principle, known as holographic Dark Energy
\cite{susskind1995world, wang2017holographic}. These models build
on the concept that the entropy associated with the boundary is
proportional to the area principle, initially proposed by
Bekenstein and Hawking in the context of black holes. According to
the Bekenstein-Hawking area law, the entropy of a black hole is
expressed as \cite{zee2001einstein}
\begin{equation}
    S_{\text{BH}} = \frac{A}{4G},
    \label{eq1}
\end{equation}
where $A$ represents the horizon area, and $G$ denotes the
gravitational constant. We also choose units in which Planck’s
constant, the speed of light, and the Boltzmann constant are set
to unity ($\hslash = c = k_B = 1$). Thus, through the lens of the
gravity-thermodynamics conjecture, we can derive the Friedmann
equations from the foundational principles of thermodynamics. It
is noteworthy that we introduce the assumption that, once
equilibrium is reached, the temperature of the Universe, along
with the cosmic horizon, is the same, which is valid in the
context of the late-time Universe. Therefore, we can comprehend
standard cosmology from an alternative perspective.

However, the crucial issue is related to resolving the crisis
faced by the standard model of cosmology. In this direction, akin
to modified gravity theories, modifying the entropy associated
with the horizon also influences the equations governing the
evolution of the universe \cite{nojiri2021different}. In fact,
there are many proposals for generalized versions of the entropy.
Taking into account systems with long-range interactions, such as
gravitational systems, which are categorized as non-extensive
systems, Tsallis introduced a generalization of standard
thermodynamics in 1988 \cite{tsallis1988possible}. This extension
broadens its applicability to non-extensive scenarios while
retaining the standard Boltzmann-Gibbs theory as a limiting case
\cite{tsallis1988possible}. It is, thus, noteworthy that, when
equipped with a specific entropy expression within any gravity
theory, it becomes possible to rephrase the Friedmann equations of
the Friedmann-Robertson-Walker (FRW) universe, which are extracted
from the first law of thermodynamics on the apparent horizon, and
vice versa. Considering this procedure, recent developments have
led to significant interest in utilizing non-extensive Tsallis
entropy for the entropy of gravitational systems
\cite{saridakis2018holographic, tavayef2018tsallis,
da2021cosmological}. This interest extends to the thermodynamic
interpretation of gravitational field equations and the
consequential effects of the non-extensive parameter within
cosmology. Notably, this can result in variations in the
gravitational constant’s strength, which consequently impacts
the energy density of the dark components of the universe. This
modification can explain the late-time acceleration of the
universe without invoking a dark energy component \cite{tavayef2018tsallis, zamora2022thermodynamically}.

Another well-known entropy, which comes from the relativistic
generalization of the standard statistical theory, is the
so-called Kaniadakis entropy \cite{kaniadakis2005statistical}.
This entropy can also be applied to black hole thermodynamics.
Using the Kaniadakis entropy, modified Friedmann equations were
derived, leading to novel physically significant terms that
contribute to an effective dark energy sector of the universe
\cite{lymperis2021modified}. While these theoretical frameworks
offer interesting possibilities for exploring the nature of dark
energy, benchmarking these models against observational data, such
as those from cosmic microwave background radiation or large-scale
structure surveys, is essential. The models need to be consistent
with the observed behavior of the universe. However, without a
reliable theory of quantum gravity, discussing the correction
terms to the entropy of a black hole in a precise manner remains
elusive.
\section{Generalized Entropy in Our Model}
\label{p3}

In this section, we propose an alternative form of entropy inspired by considerations from quantum gravity.

According to general relativity, the horizon of a black hole is precisely defined by the Schwarzschild radius. However, from a quantum mechanical perspective, understanding the geometry of spacetime requires measurements involving tools such as rods and clocks. These measurements are subject to inherent uncertainties \cite{garay1995quantum, kempf1995hilbert}. For instance, attempting to measure a distance with high precision, especially at very short scales, encounters limitations imposed by the Heisenberg uncertainty principle. Precisely determining the position of the measuring rod introduces significant uncertainty in its momentum, which in turn results in uncertainty in the rod's energy. Since energy contributes to spacetime curvature through general relativity, this energy uncertainty induces distortions in spacetime, leading to a fluctuating geometry. Thus, measurements at extremely small scales inherently face precision limits due to the interplay of quantum mechanics and general relativity.

This phenomenon persists even in the weak-energy regime of quantum gravity. Any consistent theory that fully reconciles gravity with quantum mechanics must accommodate this fundamental argument. The interplay of quantum uncertainty, energy fluctuations, and spacetime curvature highlights a fundamental limitation in measurement precision, underscoring the need for a deeper understanding of quantum spacetime.

\begin{figure}[h]
	\centering
	\includegraphics[width=2in]{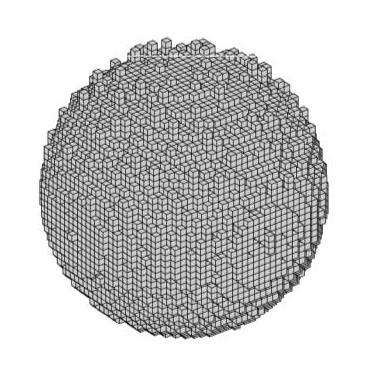}
	\caption{An exaggerated representation of a black hole's surface at Planck-scale resolution, illustrating quantum fluctuations and the fuzziness of the event horizon.}
	\label{fig1}
\end{figure}

Given the limitations on measurement precision imposed by quantum mechanics, it is reasonable to assume a degree of roughness in the surface area of a black hole when examined at the small scales, as shown in Fig.~\ref{fig1}. This roughness arises as a random distribution, potentially rooted in a quantum theory of gravity where uncertainty plays a pivotal role. Larger surface areas correspond to more fluctuations, while smaller surfaces exhibit fewer fluctuations.

Assuming that these fluctuations are homogeneously distributed, the number of fluctuations scales proportionally with the surface area. For example, doubling the surface area results in a doubling of the number of fluctuations. Based on this proportionality, the corrected surface area can be expressed as:
\begin{equation}
	\mathcal{A} = A_0 + \gamma A_0,
	\label{eqArea}
\end{equation}
where $A_0$ represents the area of the black hole as defined by its Schwarzschild radius, and the second term accounts for the surface roughness. Here, $\gamma$ is a parameter quantifying the quantum mechanical fuzziness in the black hole’s surface area. While the average volume remains unchanged, tiny ripples can significantly increase the surface area and intuitively we expect $\gamma$ is always positive. 

Building on this concept, the generalized form of black hole entropy can be written as:
\begin{equation}
	S = \frac{A_0}{4G} (1 + \gamma),
	\label{eqEntropy}
\end{equation}
where $A_0$ again denotes the horizon area. To test the validity of this model, we examine its cosmological implications by employing the thermodynamic interpretation of the gravitational field equations, modified by the generalized entropy. This leads to the derivation of modified Friedmann equations for the universe. By exploring these modifications, we can uncover new features and constraints of the model and potentially introduce innovative methods for explaining observational data in cosmology.

\section{Deriving the Modified Friedmann Equations}
Within the framework of the FRW Universe, the metric’s line
element takes the form \cite{zee2001einstein}
\begin{align}
    ds^2 = -dt^2 + a^2(t) \left[ \frac{dr^2}{1 - kr^2} + r^2 (d\theta^2 + \sin^2 \theta d\phi^2) \right],
    \label{eqFRW}
\end{align}
where $a(t)$ is the Universe’s scale factor, $k = 0, \pm 1$ denotes the curvature parameter, and $(t, r, \theta, \phi)$ are comoving coordinates. The assumption $a_0 = a(t = t_0) = 1$ is made, representing the scale factor at the present time.

Taking the apparent horizon as the Universe’s boundary, the associated temperature is determined by \cite{akbar2007thermodynamic}
\begin{align}
T_h = \frac{-1}{2 \pi r_h} \left( 1 - \frac{\dot{r_h}}{2H r_h} \right),
    \label{eqTemp}
\end{align}
where $r_h = {1}/{\sqrt{H^2 + k/a^2}}$ represents the apparent horizon radius.
Examining the apparent horizon from a thermodynamic perspective, it emerges as a suitable boundary complying with the first and second laws of thermodynamics. The energy-momentum tensor of the Universe is assumed to take the form $T_{\mu\nu} = (\rho + p)u_{\mu}u_{\nu} + p g_{\mu\nu} $, where $\rho$ denotes the energy density, and \( p \) represents the pressure. The conservation law leads to the continuity equation $\dot{\rho}+3H(\rho+p)=0$, where $ H = {\dot{a}}/{a}$ is the Hubble parameter. The work density associated with the Universe's expansion is $ W ={(\rho-p)}/{2}$ \cite{hayward1998unified}.

To apply the gravity-thermodynamics conjecture, we assert that the first law of thermodynamics on the apparent horizon holds
\begin{equation}
    dE = T_h dS + W dV.
\end{equation}
Denoting the total energy of the universe as $E = \rho V$, where $ V = \frac{4}{3} \pi r_h^3$, differentiating this expression and substituting $\dot{\rho}$ from the continuity equation, we obtain
\begin{equation}
    dE = 4\pi r_h^2 \rho dr_h - 4\pi H r_h^3 (\rho + p) dt.
\end{equation}
From the entropy relation in Eq.\ref{eqEntropy}, we find $dS=2\pi (1 + \gamma)r_h dr_h/G$. Considering the horizon temperature and applying the first law of thermodynamics,  we get
\begin{equation}
    \frac{1 + \gamma}{4\pi G r_h^3} dr_h = H (\rho + p) dt.
    \label{Eq9}
\end{equation}
Using the continuity equation and integrating both sides, we derive
\begin{equation}
    \rho = \frac{3(1 + \gamma)}{8\pi G r_h^2}.
    \label{10}
\end{equation}
Considering the relation between $r_h$ and the cosmological parameters, we can rewrite Eq.\ref{10} as
\begin{equation}
    \left(H^2 + \frac{k}{a^2}\right)=\frac{8\pi G}{3(1 + \gamma)} \rho.
    \label{mFRW1}
\end{equation}
This equation provides a generalized form of the first Friedmann equation, where the modified factor $(1 + \gamma)$ comes from the entropy correction.

The second Friedmann equation can be obtained by differentiating Eq.\ref{mFRW1} and combining it with the continuity equation. After some algebra, we arrive at
\begin{equation}
    \frac{\ddot{a}}{a}=-\frac{4\pi G}{3(1+\gamma)} (\rho+3p).
        \label{mFRW2}
\end{equation}
It is worth mentioning that although the root of the correction comes from the quantum mechanical fuzziness in the area of the black hole, it is possible to consider these effects as changing the gravitational constant, such that
\begin{equation}
    G_{\text{eff}} = \frac{G}{1 + \gamma}.
\end{equation}

\begin{figure}[h]
    \centering
    \includegraphics[width=3.5in]{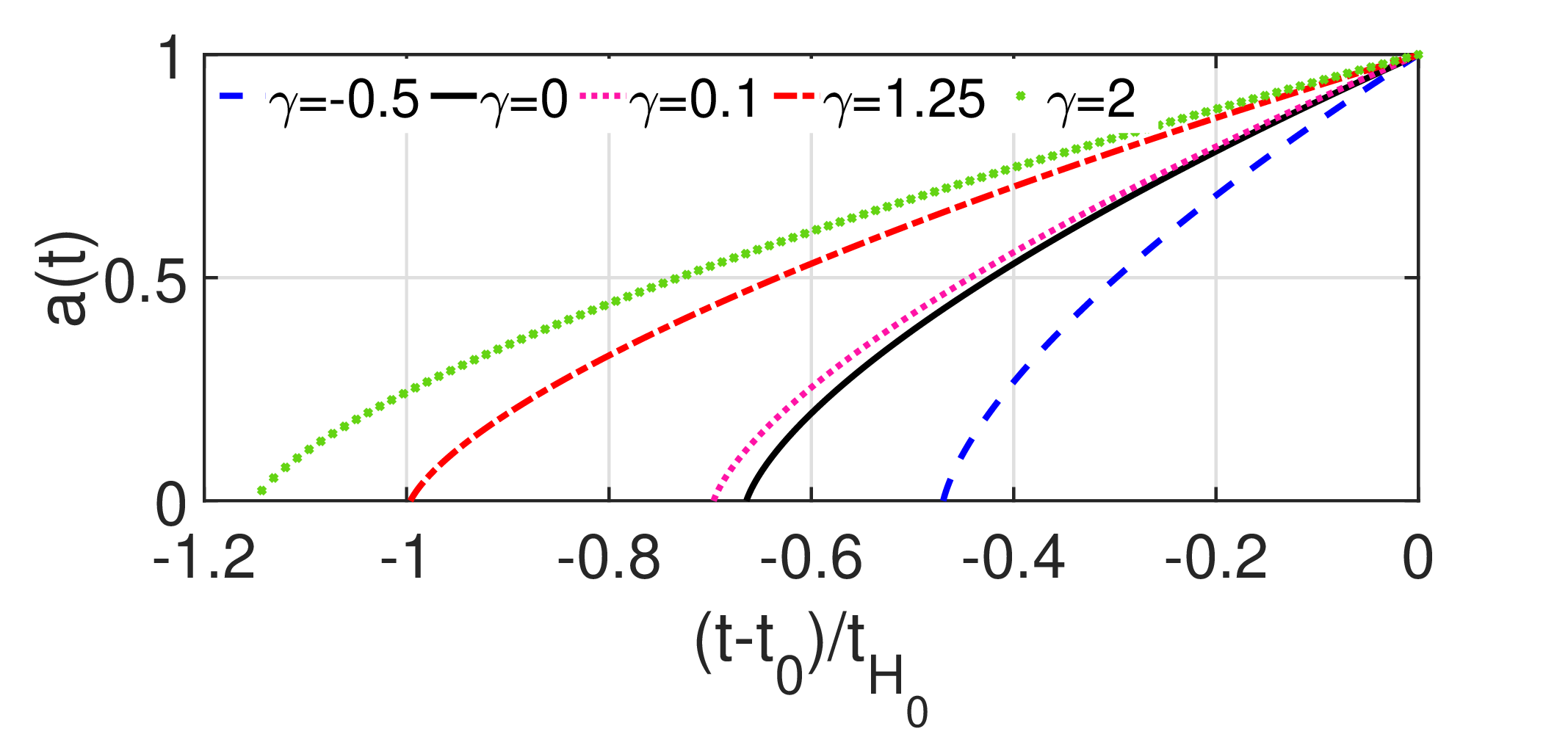}
    \caption{Evolution of the scale factor for various parameter choices of generalized entropy in a flat, matter-dominated universe. The solid line represents the scenario without corrections to the Bekenstein-Hawking entropy. The figure illustrates the effects of parameter on evolution of the universe.}
    \label{fig2}
\end{figure}

To analyze the model, we first derive the relevant cosmological parameters. For a flat, matter-dominated universe, using Eq.\ref{mFRW1}, the scale factor is expressed as
\begin{equation}
    a(t) = \left( \frac{6\pi G}{1 + \gamma} \rho_0 \right)^{1/3} t^{2/3},
    \label{at}
\end{equation}
The scale factor is modified by a factor of ${1}/{(1 +
\gamma)^{1/3}}$. In Fig.~\ref{fig2}, we illustrate the evolution
of the scale factor for a flat, matter-dominated universe under
five distinct parameter choices. The solid line represents the
scenario without any corrections to the entropy, corresponding to
the Bekenstein-Hawking entropy. In this case, the universe rapidly
enters a decelerating phase. To address this challenge, the
introduction of a dark energy component becomes necessary.

Intriguingly, altering the parameters of the generalized entropy modifies this behavior, producing results that align more closer with observational data in some range of parameters.
The age of the universe can be computed using Eq.\ref{at}, which gives
\begin{equation}
    t_{\text{age}} = \frac{2}{3} \sqrt{1 + \gamma} \, t_{H_0},
    \label{tu}
\end{equation}
Here, $t_{H_0} = 1/H_0$ denotes the Hubble time.
Deviations from the Bekenstein-Hawking entropy are accounted for through the correction parameter $\gamma$ in Eq.\ref{eqEntropy}.
As illustrated in Fig.\ref{fig2}, these adjustments have a substantial impact on the evolution of the scale factor and the universe's expansion dynamics.

From Fig.\ref{fig2}, it can be observed that for $0 < \gamma < 1.2$, the model effectively modifies standard cosmological predictions.
However, for $\gamma$ values exceeding this range, estimating the age of the universe becomes problematic.
This analysis imposes a constraint on the permissible range of the correction term $\gamma$, ensuring consistency with observational data.

To deepen our understanding of entropy in this context, we analyze the luminosity distance and its dependence on redshift, both of which are pivotal for studying cosmic dynamics.
The luminosity distance, which quantifies how an object's apparent brightness relates to its intrinsic luminosity, plays a central role in probing the universe's expansion.
It is a function of redshift ($z$), which measures the stretching of light caused by cosmic expansion.
The relationship is expressed  as \cite{weinberg2008cosmology}
\begin{equation}
    d_L(z) = (1 + z) \int_0^z \frac{dz'}{H(z')},
    \label{13}
\end{equation}
This expression encapsulates how the changing expansion rate
impacts the observed brightness of distant objects, revealing key
insights into the universe’s acceleration and the influence of
dark energy.

\begin{figure}[h]
    \includegraphics[width=3.5 in]{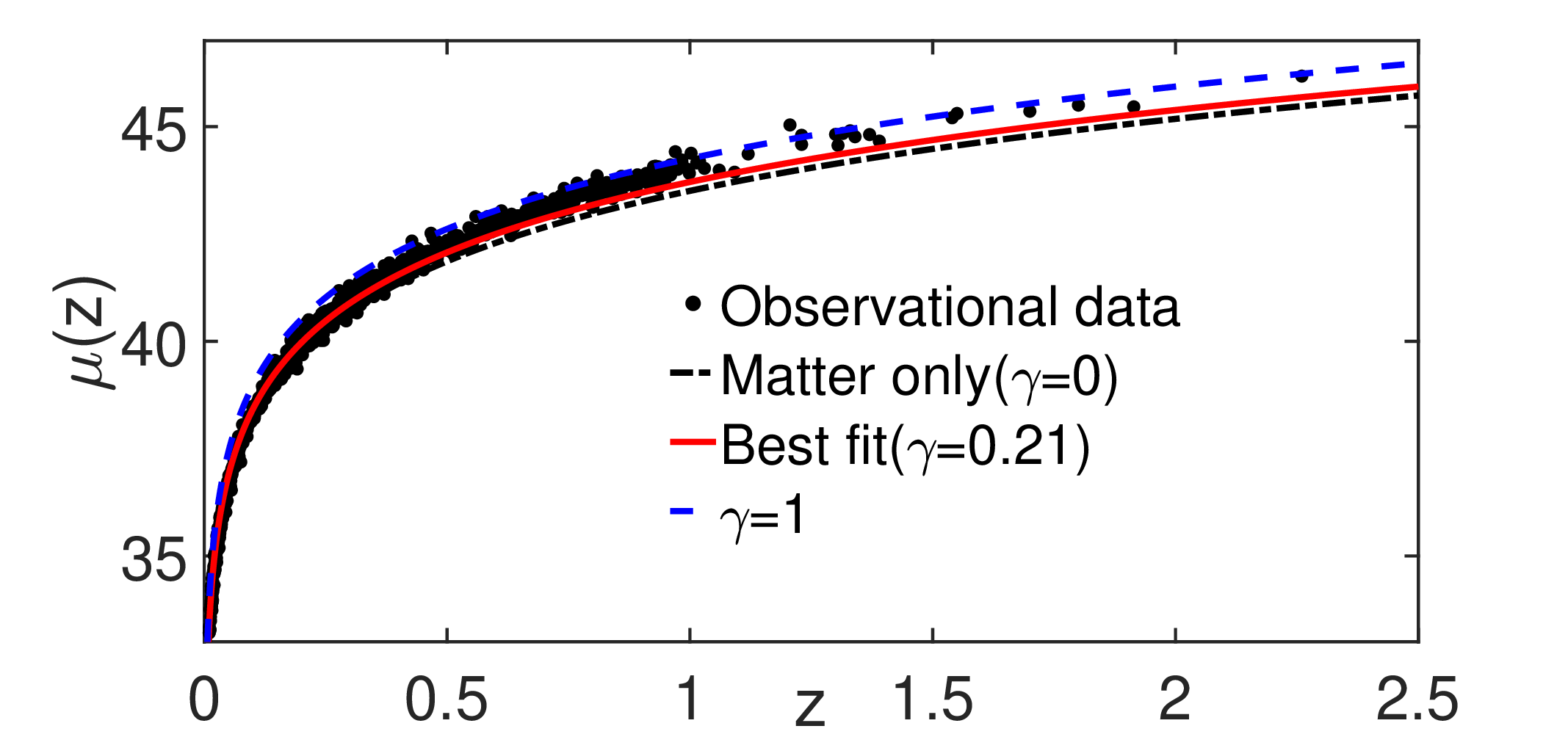}
    \centering
    \caption{Distance modulus ($\mu$) versus redshift of 1048 supernovae from the Pantheon sample \cite{scolnic2018complete}. The solid line represents the best-fit with $\gamma = 0.21$ for the modified entropy, while the dashed line corresponds to a flat matter-only universe. The supernovae clearly appear fainter (or more distant) than predicted in a matter-only universe.}
    \label{fig5}
\end{figure}

In our case, we can use the modified form of the first Friedmann equation (Eq.\ref{mFRW1}) for a flat universe and apply the scale factor-redshift relation. Substituting the Hubble parameter $H(z) = H_0 (1+z)^{3/2}/(\sqrt{1+\gamma})$, we can insert this into Eq.\ref{13}. After performing the integration, the result is
\begin{equation}
    d_L(z) = \frac{2\sqrt{1+\gamma}}{H_0} \left(1+z-\sqrt{1+z}\right)
\end{equation}
which due to the entropy modification, the supernova appears in larger distances by a factor of $\sqrt{1+\gamma}$.\\
In cosmology, the distance modulus ($\mu$) is a widely used
parameter, defined as $\mu \equiv m-M$, where $m$ represents the
apparent brightness and $M$ is the absolute brightness of an
object. Its relationship with the luminosity distance is expressed
as \cite{weinberg2008cosmology}
\begin{equation}
    \mu = 5 \log \left( \frac{d_L}{\text{Mpc}} \right) + 25,
     \label{19}
\end{equation}
where $d_L$ is the luminosity distance in Megaparsecs (Mpc). \\
In Fig.~\ref{fig5}, we present the distance modulus as a function
of redshift for 1,048 supernovae from the Pantheon sample
\cite{scolnic2018complete} and the predictions of our model in a
flat, matter-only universe. The dot-dashed line represents the
matter-only universe without considering the modified entropy. The
supernovae in the dataset appear fainter (or more distant) than
predicted by the matter-only model. The solid line represents the
best fit to the data using the modified entropy model with $\gamma
= 0.21$, while the dashed line corresponds to the case of $\gamma
= 1.0$, which diverges much from the observational data.

\begin{figure}[h]
	\includegraphics[width=3.3 in]{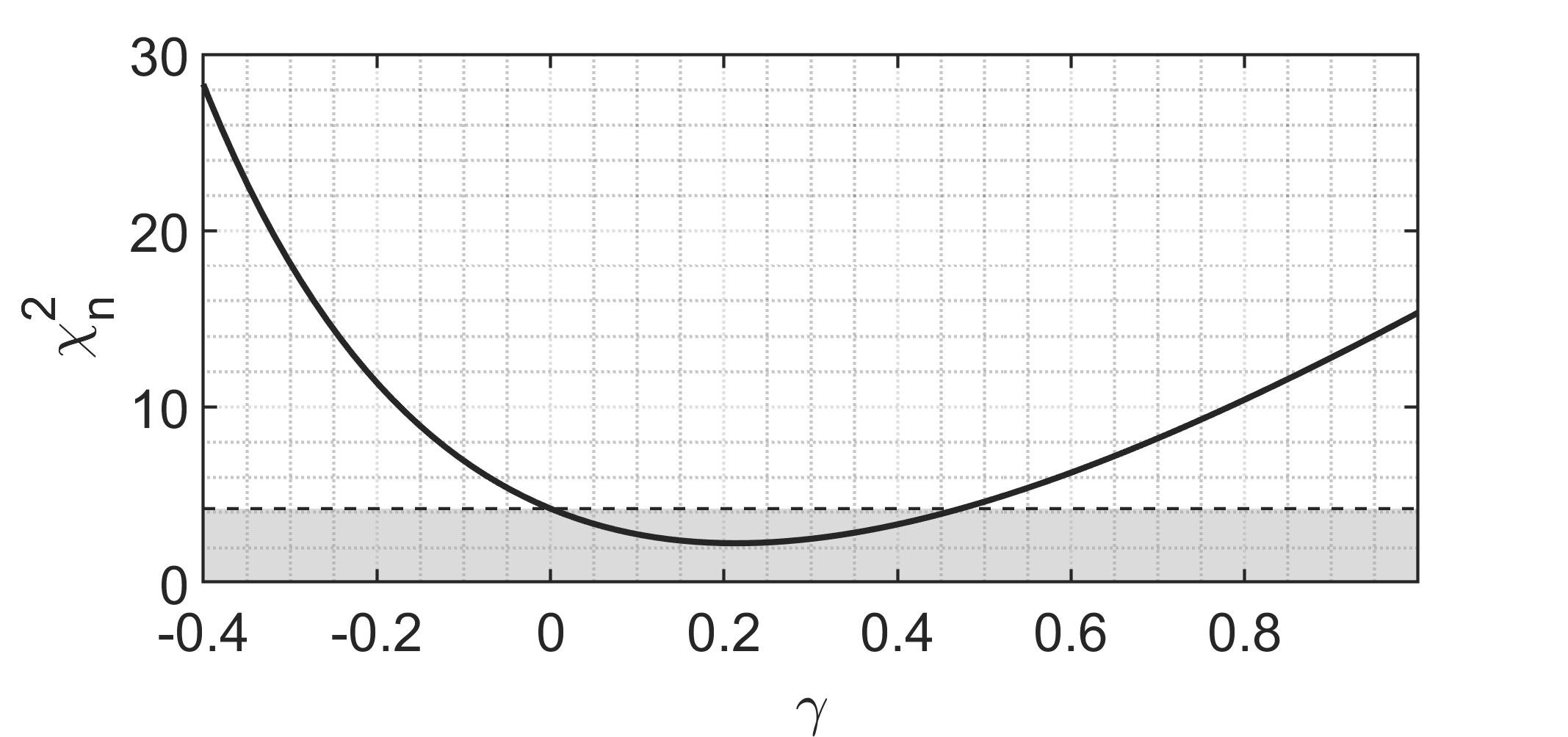}
	\centering
	\caption{Normalized $\chi^2$ ($\chi_n^2$) as a function of the roughness parameter $\gamma$. The plot highlights the improvement in the fit at $\gamma = 0.21$. The shaded gray region indicates the range where the model operates without significant observational inconsistencies.}
	\label{xi2Obs}
\end{figure}

To evaluate the deviation of the model from observational data and
determine the best fit, we employed the chi-squared ($\chi^2$)
parameter, defined as \cite{witkov2019chi}
\begin{equation}
\chi^2 = \sum_{i=1}^{N} \frac{(O_i - M_i)^2}{\sigma_i^2},
\end{equation}
Where $O_i$ is the $i$-th observed value, $M_i$ is the $i$-th
model prediction, $\sigma_i$ is the uncertainty in the $i$-th
observation and $N$ is the total number of data points. The
$\chi^2$ statistic provides a measure of how well the model fits
the data. A smaller value of $\chi^2$ indicates a better fit to
the observations. In Fig.~\ref{xi2Obs}, we plot the normalized
$\chi_n^2$, which is the $\chi^2$ value divided by the degrees of
freedom, defined as the total number of data points minus the
number of free parameters. In our case, this value is $1047$.

We observe that for $\gamma = 0.21$, the model fits the data
significantly better and can effectively account for the role of
dark energy to a good extent. However, for more precise results,
contributions from dark energy-like models, in addition to matter,
remain necessary. For instance, if we compute the deceleration
factor, it becomes $0.5$, which indicates that the matter-only
universe in this model decelerates as expected. However, due to
the weakness of gravity in this model, the universe expands more,
showing potential to reduce the gap between a matter-only universe
and observational reality. Furthermore, from the
Fig.~\ref{xi2Obs}, we deduce that for $\gamma > 0.47$, this model
becomes problematic in cosmology. Interestingly, the negative
effects of quantum fluctuations in the area of the event horizon
also present challenges, supporting the intuitive picture
discussed in Sec.~\ref{p3}. This analysis consequently imposes
constraints on $\gamma$, the quantum mechanical correction
parameter. The shaded gray region represents the parameter range
where the model performs reliably, avoiding major observational
discrepancies. As a result, the area of the event horizon can
increase by at most 47\% due to quantum fluctuations. Beyond this
threshold, new challenges emerge when attempting to reconcile
theoretical predictions with observational data.

\section{Summary and Conclusion}

We have investigated the impact of quantum corrections to the
event horizon entropy on cosmological dynamics, providing new
insights into the interplay between quantum gravity and
large-scale structure. By incorporating quantum fluctuations into
the black hole surface area, the classical Bekenstein-Hawking
entropy is generalized with a correction term proportional to a
roughness parameter, $\gamma$. According to the
thermodynamics-gravity conjecture, any correction to the entropy
leads to modifications of the Friedmann equations, effectively
introducing a $\gamma$-dependent factor that alters the evolution
of the scale factor, $a(t)$, and the universe's expansion history.\\
Our analysis reveals that small corrections produce negligible
deviations from standard cosmology, while moderate values yield
significant changes in cosmological behavior. Larger corrections
($\gamma > 0.47$) result in inconsistencies with observational
constraints, including the distance modulus versus redshift data
from the supernovae data set. This analysis of compatibility with
empirical data demonstrates that quantum fluctuations can lead to
a maximum $47\%$ increase in event horizon area without
conflicting with observations. These findings highlight the
sensitivity of the universe's evolution to quantum-gravitational
effects and offer an alternative framework to address cosmological
problems. The study provides a quantitative basis for further
exploration of quantum gravity influence on cosmology, with
observational constraints on $\gamma$ serving as a critical bridge
between theory and experiment.\\
Moreover, this model logically follows from the effects of quantum mechanics' random distributions on spacetime geometry in the low-energy limit. Consequently, any quantum gravity theory consistent with quantum mechanical foundations is expected to exhibit similar properties.\\
 Future work could refine these
constraints with different observational datasets and extend the
framework to include additional quantum-gravitational phenomena.
This investigation underlines the potential of quantum entropy
corrections to address unresolved questions in fundamental physics
and cosmology.

\bibliographystyle{apsrev}
\bibliography{library}
\end{document}